\documentclass[
superscriptaddress,
 amsmath,amssymb,
 aps,
 prl,
 10pt,
 twocolumn,
floatfix,
longbibliography
]{revtex4-2}

\usepackage[colorlinks,linkcolor=blue,anchorcolor=blue,citecolor=blue,urlcolor=blue]
{}
\usepackage{graphicx}
\usepackage{placeins}
\usepackage{color}
\usepackage[T1]{fontenc}

\begin{document}

\date{\today}

\title{Deviations from the Isobaric Multiplet Mass Equation due to Threshold States.}

\def\WUPHYS{Department of Physics, Washington University, St. Louis, Missouri 63130, USA.}
\def\FRIB{Facility for Rare Isotope Beams, Michigan State University, East Lansing, Michigan 48824, USA.}
\def\PAMSU{Department of Physics \& Astronomy, Michigan State University, East Lansing, Michigan 48824, USA.}

\def\Fudan{Key Laboratory of Nuclear Physics and Ion-beam Application (MOE), Institute of Modern Physics, Fudan University, Shanghai 200433, China.}
\def\Shanghai{Shanghai Research Center for Theoretical Nuclear Physics, NSFC and Fudan University, Shanghai 200438, China.}
\def\MSUPHYS{Department of Physics and Astronomy, Michigan State University, East Lansing, Michigan 48824, USA.}
\def\MSUCHEM{Department of Chemistry, Michigan State University, East Lansing, Michigan 48824, USA.}
\def\WUCHEM{Department of Chemistry, Washington University, St. Louis, Missouri 63130, USA.}
\def\ANL{Physics Division, Argonne National Laboratory, Argonne, IL 60439, USA.}
\def\WesternM{Department of Physics, Western Michigan University, Kalamazoo, Michigan 49008, USA.}
\def\Stores{Department of Physics, University of Connecticut, Storrs, Connecticut 06269, USA.}
\def\Lanzhou{Institute of Modern Physics, Chinese Academy of Sciences, Lanzhou 730000, China.}
\def\GANIL{Grand Acc\'{e}l\'{e}rateur National d'Ions Lourds (GANIL), CEA/DSM - CNRS/IN2P3, BP 55027, F-14076 Caen Dedex, France}

\author{R.J. Charity}
\affiliation{\WUCHEM}
\author{J. Oko\l owicz}
\affiliation{\GANIL}
\author{M. P\l oszajczak}
\affiliation{\GANIL}
\author {L.G. Sobotka}
\affiliation{\WUCHEM}
\affiliation{\WUPHYS}
\author {K.W. Brown}
\affiliation{\FRIB}
\affiliation{\MSUCHEM}

\begin{abstract}
 Recent studies have completed the $A$=16 isospin quintets for states with $J^\pi$=0$^+$ and 2$^+$. The dependence of their masses as a function of isospin projection shows evidence for deviations from quadratic behavior indicating isospin violation beyond the expectation from two-body forces. The deviation is most pronounced for the 2$^+$ states. Predictions from the shell model embedded in the continuum (SMEC) allow us to explain that this isospin violation is associated with a 
 modification of the nuclear structure due to the open-quantum-system nature of the proton-rich members of the quintet. In particular, the 0$^+$ and 2$^+$ states in $^{16}$Ne and the 2$^+$ state in $^{16}$F are threshold resonances located just above a proton-decay threshold where $s$-wave coupling to the continuum is expected.  The measured  deviations of these threshold states from the quadratic behavior of the remaining members of the multiplets  makes it possible to obtain information on the magnitude and the energy dependence of the  continuum-coupling energy correction. Continuum coupling is also indicated for the ground state of $^{8}$C, but this time through $p$-wave coupling.

\end{abstract}

\maketitle

From its introduction, the approximate symmetry captured by the concept of isospin \cite{Heisenberg:1932,Breit:1936} has been one of only a few indispensable principles that define nuclear structure \cite{Wilkinson:1969}. This strong statement is actually an understatement for light nuclei for which the Coulomb interaction has not obscured this symmetry. To the extent isospin symmetry is exact and there is no $n$-$p$ mass difference and no Coulomb interactions, then the masses of all members of an isospin multiplet would be identical. Wigner showed that when only two-body forces were responsible for the charge-dependent effects, then, in first-order perturbation theory, the masses of states with the same isospin $T$ but different isospin projections $T_{Z}\equiv(N-Z)/A$, i.e. in different nuclei, display a quadratic dependence on $T_{Z}$, 
\begin{equation}
    M(T_Z) = a + b  T_{Z} + c  T_Z^2.
\end{equation}
This dependence is called the isobaric multiplet mass equation (IMME) \cite{Wigner:1957}.  

The quadratic nature of this relationship can be tested with isospin quartets ($T=3/2$) and quintets ($T$=2) which have more than three members. In the most recent tabulation of these multiplets \cite{MacCormick:2014}, most can be well fit with a quadratic dependence. Understandably, the focus is on those with statistically significant deviations from this dependence. Such deviations, indicating isospin-violating effects, are quantified by either the $\chi^2/\nu$ of a quadratic fit or by the magnitude of the higher-order terms ($d T_{Z}^3 + e T_{Z}^4$ ...) required to achieve quality fits. 

The masses of the $A$=32 quintet have been measured with the best precision \cite{Triambak:2006, Kankainen:2010} and a small but statistical significant cubic term {$d$=$+$0.89(11)~keV \cite{MacCormick:2014} was required.  The isospin violation associated with this nonzero $d$ coefficient has been linked to isospin mixing of the intermediate member states with nearby lower-$T$ states \cite{Signoracci:2011}.  The $A$=9 quartet also requires a nonzero cubic term [$d$=$+$6.7(1.5)~keV] which has again been attributed to isospin mixing \cite{Brodeur:2012,Mukwevho:2018}. 
The $A$=53 quartet was found to have $d$=$+$39(11)~keV in Ref.~\cite{Zhang:2012} and has yet to be explained. 
The same is true for the $A$=8 quintet where a $d$ coefficient of around $+$10 keV is required \cite{Charity:2011,MacCormick:2014}. We have emphasized the sign of the $d$ coefficient as positive values indicate a suppressed mass for negative values of $T_{Z}$, i.e. those that might be proton unbound. 

As indicated above, deviations from the quadratic IMME behavior due to mixing of states with different isospin, associated with the imperfect nature of this quantum number, have received considerable attention. This Letter deals quantitatively with how quadratic IMME violations arise from a different, and up to this point neglected, kind of state mixing, mixing with the scattering continuum. 
Because of the changing thresholds for particle decays, coupling to the continuum will be different across the multiplet. To the extent that these effects are gradual across the multiplet, then they could be incorporated into the quadratic dependence and would go unrecognized.  However, as such continuum couplings depend sensitively on the proximity of the state to the threshold, they could be localized in just one, or maybe two, members of the multiplet and lead to deviations incapable of being captured by a quadratic behavior.  

The theory of near-threshold effects began with an analysis of the asymptotic behavior of the wave functions~\cite{1948PhRv...73.1002W,1951PhRv...81..412E,thomas1952analysis}. This led to the identification of Wigner cusps in the cross sections~\cite{1948PhRv...73.1002W}, and to an understanding of differences between the energies of excited states in mirror nuclei~\cite{1951PhRv...81..412E,thomas1952analysis}. Subsequent analyses in the framework of the $R$-matrix theory provided arguments for the existence of levels close to the threshold~\cite{1964PPS....84..681B}. Later, it was found that behavior analogous to the Wigner threshold phenomenon for cross sections occurs also in spectroscopic factors~\cite{2007PhRvC..75c1301M}, suggesting that the discrete and continuous aspects of the nuclear many-body problem combine at the particle-emission threshold. In Refs.~\cite{PhysRevC.72.024301,PhysRevC.77.054311}, a strong modification of the single-particle level structure of the deformed potential was found when approaching a continuum. Near the continuum, single-particle states with $\ell=0$ behave qualitatively differently than states with larger values~\cite{PhysRevC.72.024301,PhysRevLett.119.182502}.
 The appearance of clustering in near-threshold states was explained as a generic open-quantum-system phenomenon in which the near-threshold state 
shares many properties of the nearby decay channel~\cite{Okolowicz:2012,Okolowicz2013,Fernandez:2023}.

A threshold resonance can occur when the resonance sits energetically near a particle-decay threshold. If the resonance can couple to the continuum associated with this threshold channel, then modifications to the resonance wave function can occur, aligning it with the threshold channel structure~\cite{Okolowicz:2012,Okolowicz2013}. Examples of such states coupled via $s$-waves are the famous Hoyle state in $^{12}$C and the ground state of $^8$Be. In addition, threshold states can be found in  $^{11}$B \cite{Okolowicz:2020, Lopez-Saavedra:2022}, $^{14}$O\cite{Charity:2019a}, $^{13}$F \cite{Charity:2021}, and $^{15}$F \cite{Degrancey:2016,Alcindor:2022}. Coupling is strongest for the $s$-wave continuum and decreases as the total barrier (Coulomb plus centrifugal) increases~\cite{Okolowicz2013,Fernandez:2023}.

The $A$ = 16 quintet is a good place to look for such effects as theoretical arguments based on continuum coupling have suggested isospin-violating effects~\cite{Grigorenko:2002,Grigorenko:2015,Li:2025}. 
Indeed, the 0$^+$ and 2$^+$ states in $^{16}$Ne have been investigated as threshold states \cite{Okolowicz:2008,Okolowicz:2012,Okolowicz2013} that couple to the $s$-wave continuum, which could modify their structure as compared to the mirror states that are far removed from the continuum. In $^{16}$Ne, the 0$^+$ state sits just above the $p$+$^{15}$F(1/2$^+$) threshold and the 2$^+$ state sits just above the $p$+$^{15}$F(5/2$^+$) threshold. Coupling of the parent state to the continuum associated with $s$-wave protons  should enhance configurations with proton $s_{1/2}$ contributions. 
This is a unique situation as we are dealing here with two states ($0^+$ and $2^+$) in the same nucleus $^{16}$Ne and one state ($2^+$) in two nuclei $^{16}$Ne and $^{16}$F which all have $s$-wave coupling to the decay channel. This provides information about the dependence of continuum-coupling energy correction ($E_{\rm{corr}}$) both on the continuum-coupling strength and the energy separation from the proton-emission threshold.

In a companion paper \cite{CharityLong:2025}, measurements of $T$=2 states in $^{16}$F are presented which allow the $A$=16 isospin quintets to be completed for both the 0$^+$ and 2$^+$ sequences.  These $^{16}$F states were created following proton knockout from a $^{17}$Ne secondary beam using invariant-mass spectroscopy. The 0$^+$, $T$=2 state, the isobaric analog of $^{16}$Ne, was observed in the 2$p$+$^{14}$N(g.s.) channel. This peak was in fact observed in a previous analysis of these data, but was shifted up in energy due to less-accurate energy calibrations, making its assignment as the isobaric analog state seem unreasonable. Newer more-accurate calibrations are based on reproducing known $p$+$^{14}$N resonances in $^{15}$O. The 2$^+$, $T$=2 state was also observed in an isospin-conserving decay branch, 2$p$+$^{14}$N(0$^+$, $T$=1).

This reanalysis \cite{CharityLong:2025} of the $^{17}$Ne beam data also studied the states in $^{16}$Ne produced from neutron knockout reactions. With the large statistics obtained for the 2$p$+$^{14}$O(g.s.) channel, it was possible to select events where the heavy residue recoiled perpendicular to the beam axis. These selected events have reduced background, improved resolution, and less sensitivity to the energy calibrations allowing their energies to be extracted with higher precision. The mass excesses of these states and the corresponding $^{16}$O, $^{16}$N, and $^{16}$C values are listed in Table \ref{tbl:A=16}. The latter values are obtained from the ENSDF \cite{ENSDF} and AME2020 \cite{AME2020} databases with the exception for the 2$^+$ state in $^{16}$C, where an average of two more recent high-precision measurements \cite{Wiedeking:2008,Petri:2012} is used.  

Quadratic IMME fits to the 0$^+$ and 2$^+$ quintets were not very satisfactory with $\chi/\nu$= 2.89 and 4.08, respectively \cite{CharityLong:2025} with the deviation from quadratic behavior most pronounced for the 2$^+$ quintet. Fits with a cubic term added to the IMME reduce the fitted $\chi^2/\nu$ values to near unity, giving $d$= $+$4.0(22) and $+$8.9(31)~keV, respectively. The latter is a 2.9-$\sigma$ deviation which is basically at the threshold at which nonquadratic behavior was considered in the most recent evaluation~\cite{MacCormick:2014}. The values of the cubic terms are similar in magnitude to those extracted for the $A$=9 quartet [$d$=$+$6.7(15) keV] and the $A$=8 quintet ($\approx +$ 10 keV).

\begin{table}[ht]
\caption{Mass excesses in MeV for the 0$^+$ and 2$^+$  quintets ($A$=16) used in the IMME analysis. When known separately, the uncertainties are given by statistical and systematic values added in quadrature.} 
\label{tbl:A=16}
\begin{ruledtabular}
\begin{tabular}{ccc}
Nucleus  &  0$^+$  & 2$^+$ \\
\hline
$^{16}$Ne &   24.047(10)  &   25.766(12)           \\
$^{16}$F  &     20.821(11) &  22.588(26)            \\
$^{16}$O  &    17.984(3) & 19.785(11)  \\
$^{16}$N  &   15.612(7) & 17.385(7)   \\
$^{16}$C  &   13.69413(35) &  15.454(4)     \\
\end{tabular}

\end{ruledtabular}
\end{table}


Figure~\ref{fig:thres_16} shows the energy $E_p$ above threshold for proton emissions to the 1/2$^+$ (5/2$^+$) state in the appropriate $T$=3/2, $A$=15 daughter for the 0$^+$ (2$^+$) quintet. For the 0$^+$ quintet (red points and line), the threshold is diffuse due to the large widths of the $J^\pi$=1/2$^+$, $T$=3/2 states defining the threshold and thus there is some uncertainty in assigning the energy above threshold.

\begin{figure}[!htb]
\includegraphics[width=1.\linewidth]{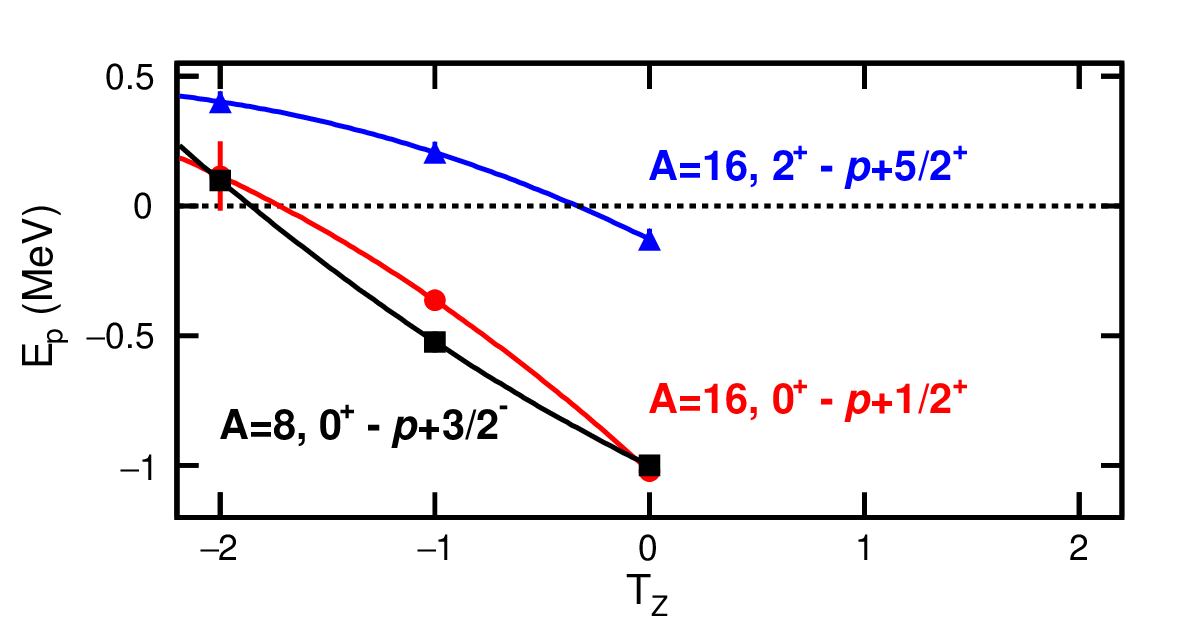}
\caption{Energy $E_{p}$ above the indicated threshold for the $A$=16 and $A$=8 quintet states as a function of their isospin projection.} 
\label{fig:thres_16}
\end{figure} 

For this mass region, the coupling to the proton $s$-wave continuum is expected to be maximized when the state is $\approx$0.55~MeV above the respective threshold, discussed below and see \cite{Okolowicz:2012,Okolowicz2013}.  From Fig.~\ref{fig:thres_16}, we therefore expect the 0$^+$ and 2$^+$ states of $^{16}$Ne and the 2$^+$ state of $^{16}$F to have  continuum $s$-wave couplings modifying their structure and breaking isospin symmetry with the largest, and most statistically quantifiable, effect for the 2$^+$ state in $^{16}$Ne.  

\begin{figure}[!htb]
\includegraphics[width=1.\linewidth]{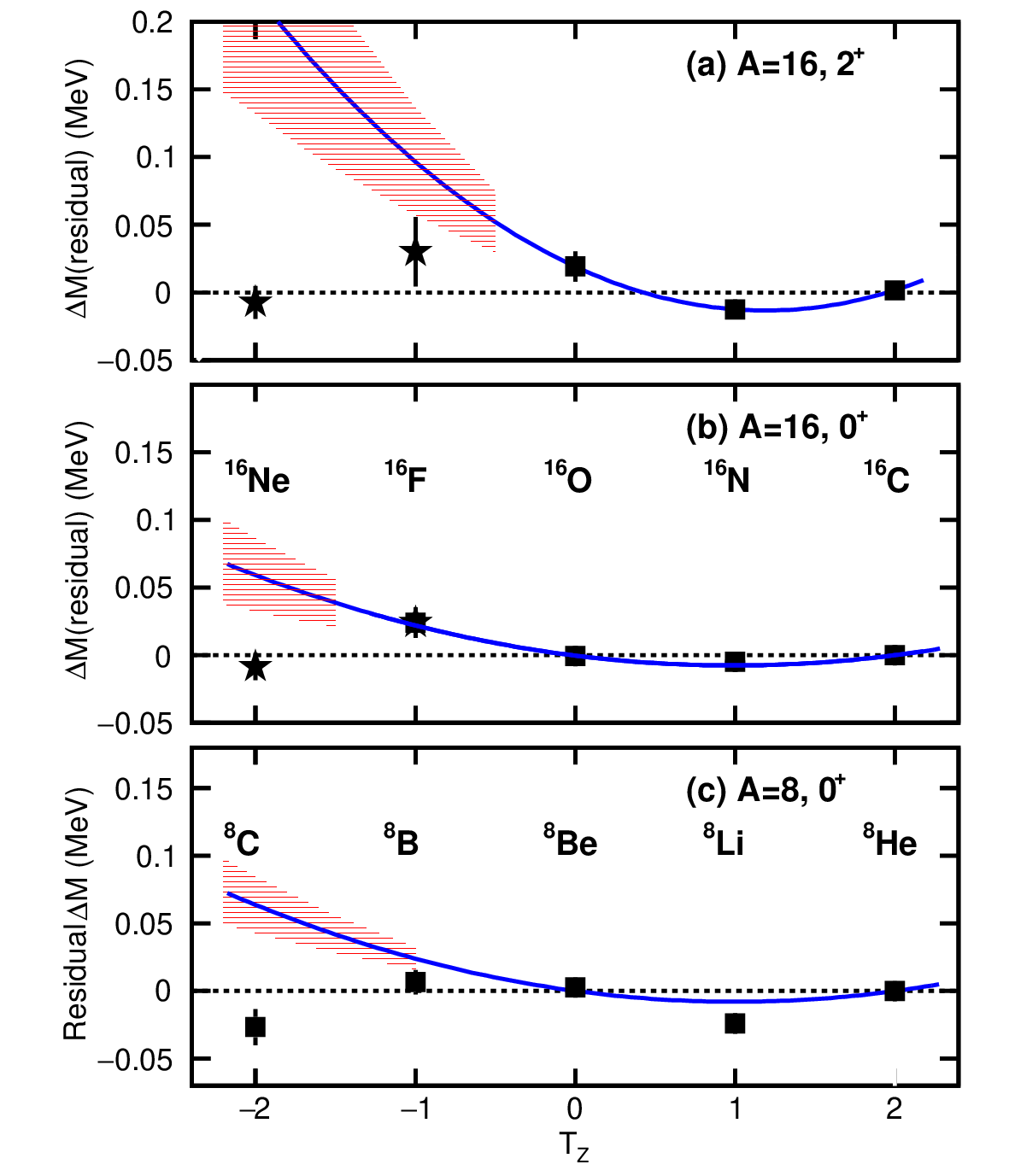}
\caption{Residuals of quadratic fits to the mass excesses of the (a) 2$^+$ and (b) 0$^+$ quintets for $A$=16 and the (c) 0$^+$ quintet for $A$=8.  The solid blue curve shows a quadratic extrapolation using the masses of neutron-rich members to estimate the values of the  proton-rich masses when threshold effects are absent. The hatched area show the $\pm$1$\sigma$ region of these extrapolated values. The stars indicate the values extracted using the new masses presented in the companion paper \cite{CharityLong:2025}.} 
\label{fig:imme_16}
\end{figure}

We can estimate the correlation energy $E_{\rm{corr}}$, the reduction in energy due to continuum coupling, from Fig.~\ref{fig:imme_16} where residuals from the quadratic fit for each quintet are plotted. In Fig.~\ref{fig:imme_16}(a) for the 2$^+$ states, the three neutron-rich members are not threshold states and have minimal effects from continuum coupling. Assuming there are no other isospin breaking effects such as isospin mixing modifying their energies, they should follow a quadratic IMME. Therefore, a quadratic extrapolation of these points to lower $T_{Z}$ can then be used to estimate what the values of the  proton-rich masses would be if continuum coupling were absent. The same can be said for 0$^+$ states shown in Fig.~\ref{fig:imme_16}(b); however, here the $^{16}$F state can be added but, with its larger uncertainty, adds little value to the quadratic extrapolation. The difference between these extrapolations (red shaded regions give the $\pm$1$\sigma$ extrapolation limits) and the experimental residuals provides a measure of $E_{\rm{corr}}$. 

A similar analysis was performed for the $A$=8 quintet in Fig.~\ref{fig:imme_16}(c) which is also known to require the presence of cubic term.  Here we use the masses from Ref.~\cite{MacCormick:2014}, except for $^8$C where we used the value $\Delta M$=35.067(13)~MeV, the weighted average of the AME2020 mass \cite{AME2020} and a newer measurement \cite{Charity:2023b}.
This quintet has one threshold state, $^8$C, which is located just above the $p$+$^{7}$B threshold (Fig.~\ref{fig:thres_16}), so the extrapolation proceeds in the same manner as the 0$^+$ A=16 case.

{\it Method, model space, and parameters---}
The continuum-coupling correlation energy $E_{\rm{corr}}$ can be calculated in the shell model embedded in the continuum (SMEC), the open quantum system realization of the standard nuclear shell model~\cite{Okolowicz:2003}. In the simplest version of the multichannel SMEC, the Hilbert space is divided into two orthogonal subspaces ${\cal Q}_{0}$ and  ${\cal Q}_{1}$ containing 0 and 1 particle in the scattering continuum, respectively. In the open-quantum-system framework, ${\cal Q}_0$  includes couplings to the environment of decay channels through the energy-dependent effective Hamiltonian:
\begin{equation}
{\cal H}(E)=H_{{\cal Q}_0{\cal Q}_0}+W_{{\cal Q}_0{\cal Q}_0}(E),
\label{eq21}
\end{equation}
where $H_{{\cal Q}_0{\cal Q}_0}$ denotes the standard shell-model Hamiltonian describing the internal dynamics in the closed quantum system approximation, and 
\begin{equation}
W_{{\cal Q}_0{\cal Q}_0}(E)=H_{{\cal Q}_0{\cal Q}_1}G_{{\cal Q}_1}^{(+)}(E)H_{{\cal Q}_1{\cal Q}_0},
\label{eqop4}
\end{equation}
is the energy-dependent continuum-coupling term, where $E$ is a scattering energy, $G_{{\cal Q}_1}^{(+)}(E)$ is the one-nucleon Green's function, and ${H}_{{Q}_0,{Q}_1}$ and ${H}_{{Q}_1{Q}_0}$ couple ${\cal Q}_{0}$ with ${\cal Q}_{1}$. 
The energy scale in Eq.~(\ref{eq21}) is defined by the nearest one-nucleon emission threshold. The channel state is defined by the coupling of one nucleon in the scattering continuum to a shell-model wave function of the $(A-1)$ nucleus. 

The continuum-coupling correlation energy $E_{\rm{corr}}$ in the eigenstate $|\Psi_{\alpha}^{J^{\pi}}\rangle$ of ${\cal H}(E)$ is~\cite{Okolowicz:2012,Okolowicz2013}:
 \begin{equation}
E_{{\rm corr};\alpha}^{J^{\pi}}(E)=\langle \Psi_{\alpha}^{J^{\pi}}(E)| {\cal H} |\Psi_{\alpha}^{J^{\pi}}(E)\rangle - \langle \Phi_i^{J^{\pi}}|H_{{\cal Q}_0{\cal Q}_0}|\Phi_i^{J^{\pi}} \rangle.
\label{eq22}
\end{equation}
The SMEC eigenstate $|\Psi_{\alpha}^{J^{\pi}}\rangle$ is the linear combination of shell-model eigenstates $\{|\Phi_{j}^{J^{\pi}}\rangle\}$ and $|\Phi_{i}^{J^{\pi}}\rangle$ in Eq.~(\ref{eq22}) 
is the shell-model eigenstate with largest weight in the decomposition of the SMEC eigenstate $|\Psi_{\alpha}^{J^{\pi}}\rangle = \sum_j b_{\alpha j}^{J^{\pi}} |\Phi_{j}^{J^{\pi}}\rangle$.
 For a given total angular momentum $J$ and parity ${\pi}$, the mixing of shell-model states in a given SMEC eigenstate is due to their coupling to the same decay channel $(\ell j)$. In Eq.~(\ref{eq22}), for a given energy $E$, one selects the eigenstate $|\Psi_{\alpha}^{J^{\pi}}(E)\rangle$, which has the correct one-nucleon asymptotic behavior. For that, the depth of the average potential is chosen to yield the single-particle energy  equal to  $E$.  The point of the strongest collectivization corresponds to the energy at which the continuum-coupling correlation energy is minimal and is determined by the interplay between Coulomb-plus-centrifugal barrier and the continuum coupling.

For $H_{{\cal Q}_0{\cal Q}_0}$ we take Zuker-Buck-McGrory~\cite{ZBM,ZBM1} and Cohen-Kurath~\cite{Cohen-Kurath} effective Hamiltonians for $A=16$ and $A=8$~\cite{Shyam2000}, respectively.
The continuum-coupling interaction is given by the Wigner contact force with a coupling strength $|V_{12}^{(0)}|$=510 MeV fm$^3$ and 650 MeV fm$^3$ for $A=16$~ and $A=8$, respectively~\cite{Shyam2000}. The radial single-particle wave functions (in ${\cal Q}_0$) and the scattering wave functions (in ${\cal Q}_1$) are generated by the average potential which includes the central Woods-Saxon term, the spin-orbit term, and the Coulomb potential. The radius and diffuseness of the Woods-Saxon and spin-orbit potentials are $R_0=1.27 A^{1/3}$~fm and $a=0.67$~fm, respectively. The strength of the spin-orbit potential is $V_{\rm SO}=5.78$ MeV. The Coulomb part is calculated for a uniformly charged sphere with the radius $R_0$.

The continuum-coupling correlation energy is negative, i.e., the continuum coupling always lowers the energy. Coupling to the continuum will modify the energy of all members of the quintet to some extent, even $^{16}$C which is particle bound. Again, if these lowerings are gradual across the quintet, it may not lead to significant deviations from quadratic behavior. However, in the event that threshold resonances exist in one or two members of the quintet, deviations from quadratic behavior may be conspicuous. Specifically, the SMEC predicts an extra decrease in the energy for proton resonances located just above threshold, i.e., $E_{\rm{corr}}$ becomes more negative.  It is this extra decrease that could lead to a nonquadratic deviation from the IMME and, in particular, positive $d$ coefficients. 

 \begin{figure}[!htb]
 \centering
 \begin{tabular}{c}
     \includegraphics[width=.9\linewidth, clip, trim = 0 80 10 35]{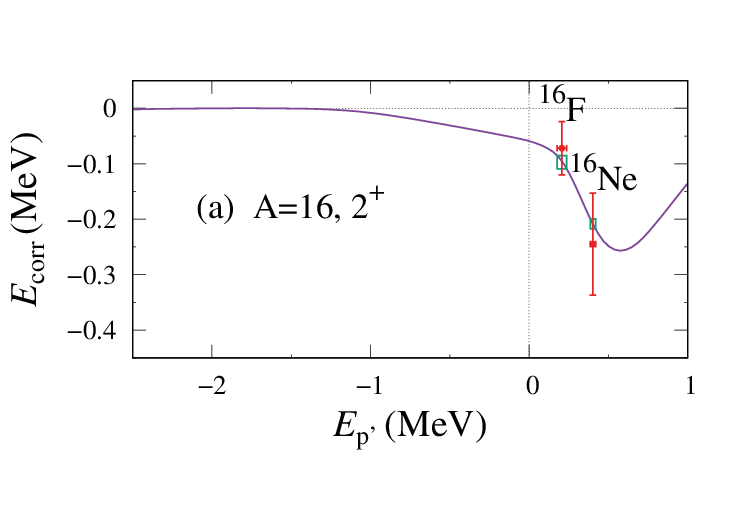}\\
     \includegraphics[width=.9\linewidth, clip, trim = 0 80 10 35]{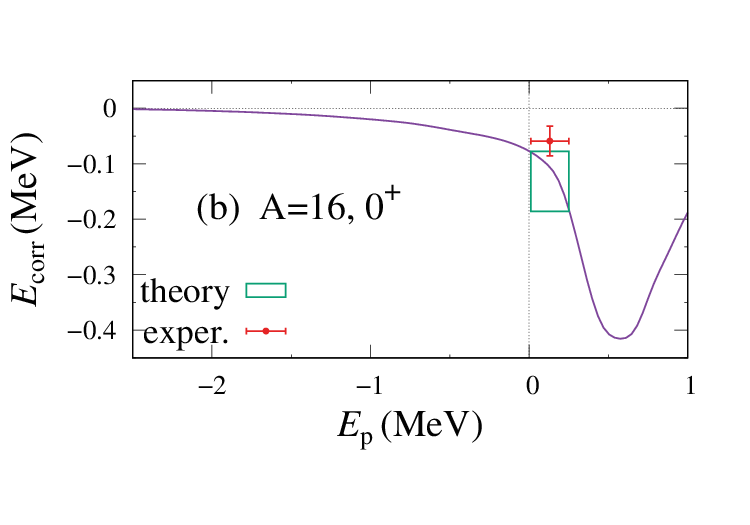} \\
     \includegraphics[width=.9\linewidth, clip, trim = 0 35 10 35]{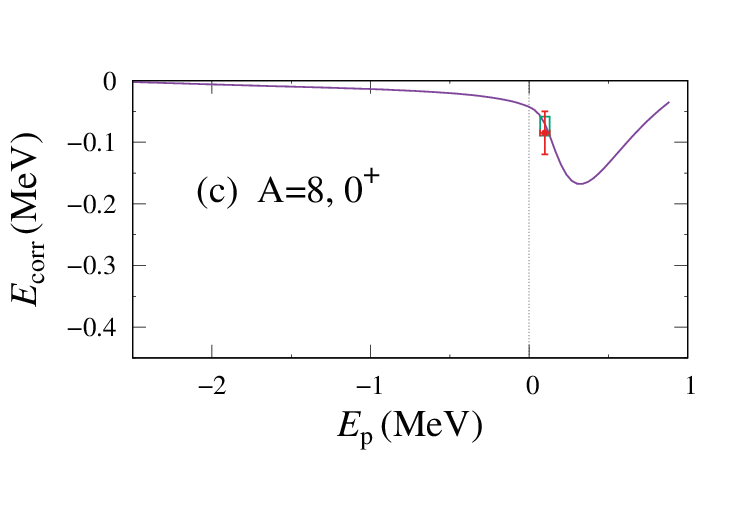}
 \end{tabular}
\caption{Magenta curves shows the predicted continuum-coupling correlation energy as a function of the energy above the appropriate threshold for the (a) 2$^+$ resonances in $^{16}$Ne and $^{16}$F, (b) 0$^+$ ground state of $^{16}$Ne, and (c) 0$^+$ ground state of ${^8}$C. The data points give the experimental values, and SMEC predictions for each data point are shown as the green squares.} 
\label{fig:SMEC}
\end{figure} 

{\it Results---}
The $E_{\rm{corr}}$ values for the threshold states in $^{16}$Ne, $^{16}$F, and $^8$C are plotted in Fig.~\ref{fig:SMEC} versus $E_p$. The theoretical uncertainties, illustrated by the green boxes, are due to the uncertainties of the experimental separation energies. For both the $0^+$ in $^{16}$Ne and the $2^+$ resonances in $^{16}$Ne and $^{16}$F, we are dealing with $\ell=0$ 
coupling to the nearest proton-decay channel. The $0^+$ resonance in $^8$C is coupled to the $\ell = 1$ proton-decay channel $p + ^7{\rm B}(3/2^-)$.  The abscissa values of the minima of the magenta curves in Fig.~\ref{fig:SMEC} indicate the optimal energy at which the magnitude of the correlation energy is largest for the considered SMEC eigenstate. These values are $E_{p} = 0.58$ MeV for $A=16$ case, and $E_{p} = 0.32$ MeV for $A=8$. These optimal energies are shifted above the threshold as a result of the interplay between the attractive coupling to the proton-decay channel, which is strongest at the threshold, and the repulsive Coulomb and centrifugal interactions. With increasing proton number, the optimal energy is shifted to higher values and the corresponding $E_{\rm{corr}}$ decreases significantly. $E_{\rm{corr}}$ becomes insignificant in nuclei above the $(sd)$ shell~\cite{Okolowicz2013}. 

The continuum-coupling energy correction $E_{\rm{corr}}$ for the $0^+$ resonance in $^8$C, $^{16}$Ne and the $2^+$ resonance in $^{16}$F are far from the value at the optimal energy. This is not the case for the $2^+$ resonance in $^{16}$Ne, where $E_{\rm{corr}}$ is close to its optimal value. One should notice that for $E_p$ higher than the optimal energy, the continuum-coupling correlation energy decreases rapidly. As a result, the range in resonance energy, above the particle-emission threshold, where coupling to the continuum can cause significant deviations from the quadratic form of the IMME is relatively narrow.


The SMEC spectroscopic factors in the considered states of $^{16}$Ne and $^{16}$F are ${\cal S}_{s1/2} = 0.41$ in $0^+$ resonances, and ${\cal S}_{s1/2} = 0.35$ in $2^+$ resonances. In $^8$C, the spectroscopic factor for $0^+$ resonance is ${\cal S}_{p3/2} = 0.75$. For these relatively large spectroscopic factors, the calculated correlation energy $E_{\rm{corr}}$ is of sufficient magnitude to shift a resonance energy enough to generate a statistically significant deviation from the quadratic IMME form.

Near-threshold variations of the spectroscopic factor are closely related with the magnitude of the $E_{\rm{corr}}$. In the studied cases of $^8$C, $^{16}$Ne, and $^{16}$F, the continuum-coupling correlation energy is small. Consequently, the variation of the spectroscopic factor with energy is small as well. For example in $^{16}$Ne($0^+$), the ratio of spectroscopic factor in the SMEC to that in the shell model at the experimental resonance energy is only 1.01, while at the optimal energy this ratio is 1.1.

{\it Conclusions---}
Deviations from the quadratic isobaric multiplet mass equation observed for the 0$^+$ ($A$=8, 16) and 2$^+$ ($A$=16) 
$T$=2 multiplets can be explained based on the changing role of the continuum across these multiplets. In particular, the 0$^+$ and 2$^+$ states in $^{16}$Ne, the 0$^+$ state in $^8$C, and the 2$^+$ state in $^{16}$F are threshold states located in energy just above a proton-decay threshold. These states have reduced  mass as inferred from quadratic extrapolations of the well-subthreshold members of each multiplet and the reductions are consistent with those predicted from a model that incorporates coupling to the continuum. Simultaneous description of the continuum-coupling correlation energy $E_{\rm{corr}}$ using the same coupling strength in the identified threshold states allows us to obtain, for the first time, information on both the magnitude of $E_{\rm{corr}}$ and its dependence  on the energy.

The magnitude of $E_{\rm{corr}}$ rises to a maximum somewhat above the decay threshold and then rapidly decreases (Fig. \ref{fig:SMEC}). Therefore only states in a narrow window above the threshold 
will give rise to these continuum effects.
Also, these effects are expected to decrease with increasing Coulomb barrier and hence, are basically restricted to nuclei in the $p$- and $sd$-shells~\cite{Okolowicz2013}. This observation is consistent with a variation of the coefficient of the cubic term added to the IMME, which changes from $d\approx +10$ keV for the A=8 quintuplet, to $d\approx +4$ keV for the A=16 quintuplet, and to $d\approx +0.9$ keV for the A=32 quintuplet. 





 \vskip 0.5truecm
 \begin{acknowledgments}
 This material is based upon work supported by the U.S.
Department of Energy, Office of Science, Office of Nuclear
Physics under Awards No. DE-FG02-87ER-40316. We gratefully acknowledge a support from the CRIANN (Normandy, France) for providing computing and data-processing resources needed for this work.

  \end{acknowledgments}


\bibliography{library}

\end{document}